\documentclass[12pt]{article}
\usepackage{epsfig}

\usepackage{mathrsfs}
\usepackage[T1]{fontenc}
\usepackage{mathpazo}
\usepackage{setspace}
\usepackage{amsfonts}
\usepackage{amssymb}
\usepackage{amsmath}
\usepackage{epsfig}
\usepackage{latexsym}
\usepackage{color}
\usepackage{graphicx}
\usepackage{nicefrac}
\usepackage[latin1]{inputenc}
\usepackage{pstricks}
\usepackage{slashed}

\def\hybrid{\topmargin -20pt    \oddsidemargin 0pt
        \headheight 0pt \headsep 0pt
        \textwidth 6.25in       
        \textheight 9.5in       
        \marginparwidth .875in
        \parskip 5pt plus 1pt   \jot = 1.5ex}

\hybrid

\catcode`\@=11

\def\marginnote#1{}
\newcount\hour
\newcount\minute
\newtoks\amorpm
\hour=\time\divide\hour by60 \minute=\time{\multiply\hour by60
\global\advance\minute by-\hour}
\edef\standardtime{{\ifnum\hour<12 \global\amorpm={am}%
        \else\global\amorpm={pm}\advance\hour by-12 \fi
        \ifnum\hour=0 \hour=12 \fi
        \number\hour:\ifnum\minute<10 0\fi\number\minute\the\amorpm}}
\edef\militarytime{\number\hour:\ifnum\minute<10
0\fi\number\minute}

\def\draftlabel#1{{\@bsphack\if@filesw {\let\thepage\relax
   \xdef\@gtempa{\write\@auxout{\string
      \newlabel{#1}{{\@currentlabel}{\thepage}}}}}\@gtempa
   \if@nobreak \ifvmode\nobreak\fi\fi\fi\@esphack}
        \gdef\@eqnlabel{#1}}
\def\@eqnlabel{}
\def\@vacuum{}
\def\draftmarginnote#1{\marginpar{\raggedright\scriptsize\tt#1}}

\def\draft{\oddsidemargin -.5truein
        \def\@oddfoot{\sl preliminary draft \hfil
        \rm\thepage\hfil\sl\today\quad\militarytime}
        \let\@evenfoot\@oddfoot \overfullrule 3pt
        \let\label=\draftlabel
        \let\marginnote=\draftmarginnote
   \def\@eqnnum{(\theequation)\rlap{\kern\marginparsep\tt\@eqnlabel}%
\global\let\@eqnlabel\@vacuum}  }

\def\preprint{\twocolumn\sloppy\flushbottom\parindent 2em
        \leftmargini 2em\leftmarginv .5em\leftmarginvi .5em
        \oddsidemargin -.5in    \evensidemargin -.5in
        \columnsep .4in \footheight 0pt
        \textwidth 10.in        \topmargin  -.4in
        \headheight 12pt \topskip .4in
        \textheight 6.9in \footskip 0pt
        \def\@oddhead{\thepage\hfil\addtocounter{page}{1}\thepage}
        \let\@evenhead\@oddhead \def\@oddfoot{} \def\@evenfoot{} }

\def\numberbysection{\@addtoreset{equation}{section}
        \def\theequation{\thesection.\arabic{equation}}}

\def\underline#1{\relax\ifmmode\@@underline#1\else
        $\@@underline{\hbox{#1}}$\relax\fi}

\def\titlepage{\@restonecolfalse\if@twocolumn\@restonecoltrue\onecolumn
     \else \newpage \fi \thispagestyle{empty}\c@page\z@
        \def\thefootnote{\fnsymbol{footnote}} }

\def\endtitlepage{\if@restonecol\twocolumn \else \newpage \fi
        \def\thefootnote{\arabic{footnote}}
        \setcounter{footnote}{0}}  

\catcode`@=12 \relax

\def\figcap{\section*{Figure Captions\markboth
        {FIGURECAPTIONS}{FIGURECAPTIONS}}\list
        {Figure \arabic{enumi}:\hfill}{\settowidth\labelwidth{Figure
999:}
        \leftmargin\labelwidth
        \advance\leftmargin\labelsep\usecounter{enumi}}}
 \relax
\def\tablecap{\section*{Table Captions\markboth
        {TABLECAPTIONS}{TABLECAPTIONS}}\list
        {Table \arabic{enumi}:\hfill}{\settowidth\labelwidth{Table
999:}
        \leftmargin\labelwidth
        \advance\leftmargin\labelsep\usecounter{enumi}}}
 \relax
\def\reflist{\section*{References\markboth
        {REFLIST}{REFLIST}}\list
        {[\arabic{enumi}]\hfill}{\settowidth\labelwidth{[999]}
        \leftmargin\labelwidth
        \advance\leftmargin\labelsep\usecounter{enumi}}}
 \relax
\makeatletter
\newcounter{pubctr}
\def\publist{\@ifnextchar[{\@publist}{\@@publist}}
\def\@publist[#1]{\list
        {[\arabic{pubctr}]\hfill}{\settowidth\labelwidth{[999]}
        \leftmargin\labelwidth
        \advance\leftmargin\labelsep
        \@nmbrlisttrue\def\@listctr{pubctr}
        \setcounter{pubctr}{#1}\addtocounter{pubctr}{-1}}}
\def\@@publist{\list
        {[\arabic{pubctr}]\hfill}{\settowidth\labelwidth{[999]}
        \leftmargin\labelwidth
        \advance\leftmargin\labelsep
        \@nmbrlisttrue\def\@listctr{pubctr}}}
 \relax
\makeatother
\newskip\humongous \humongous=0pt plus 1000pt minus 1000pt

\newif\ifdtup

\relax

\def\be{\begin{equation}}
\def\ee{\end{equation}}
\def\ba{\begin{eqnarray}}
\def\ea{\end{eqnarray}}

\renewcommand{\theequation}{\thesection.\arabic{equation}}

\newcommand{\eqn}[1]{(\ref{#1})}

\author{
  \begin{minipage}{.97\linewidth}
    \vspace{0cm}
    \begin{center}
      \begin{small}
        \textbf{Ioannis Bakas}\footnote{bakas@ajax.physics.upatras.gr} ${\ }^1$,
        \textbf{Francois Bourliot}\footnote{bourliot@cpht.polytechnique.fr} ${\ }^2$,
        \textbf{Dieter L\"ust}\footnote{dieter.luest@lmu.de} ${\ }^{3,4}$ and
         \textbf{Marios Petropoulos}\footnote{marios@cpht.polytechnique.fr} ${\ }^2$
      \end{small}
    \end{center}
    \vspace{0.5cm}
    \hspace{2cm}\begin{minipage}{.7\linewidth}
     {\it \begin{footnotesize}
    \begin{itemize}
        \item[${}^1$] Department of Physics, University of Patras\\
       26500 Patras, Greece
      \item[${}^2$] Centre de Physique Th\'eorique,
        Ecole Polytechnique\\
        CNRS UMR 7644, 91128 Palaiseau Cedex, France
      \item[${}^3$] Max-Planck-Institut f\"ur Physik\\
       F\"ohringer Ring 6, 80805 M\"unchen, Germany
                    \item[${}^4$] Arnold-Sommerfeld-Center f\"ur Theoretische Physik\\
        Department f\"ur Physik, Ludwig-Maximilians-Universit\"at M\"unchen\\
        Theresienstra\ss e 37, 80333 M\"unchen, Germany
       \end{itemize}
     \end{footnotesize}}
    \end{minipage}
    \vspace{0.5cm}
  \end{minipage}
}

\title{\vspace{0.5cm}
 \boldmath \begin{LARGE}
    \textbf{Mixmaster universe in Ho\v{r}ava-Lifshitz gravity}
  \end{LARGE} \unboldmath
}

\begin{document}

\renewcommand{\thepage}{\arabic{page}}
\setcounter{page}{1}


\begin{titlepage}
  \maketitle
  \thispagestyle{empty}

  \vspace{-14.2cm}
  \begin{flushright}
    CPHT-RR111.1109\\
    LMU-ASC 48/09\\
    MPP-2009-180
  \end{flushright}

  \vspace{11cm}

  \begin{center}
    \textsc{Abstract}\\
  \end{center}
  We consider spatially homogeneous (but generally non-isotropic) cosmologies in
  the recently proposed Ho\v{r}ava-Lifshitz gravity and compare them to those of
  general relativity using Hamiltonian methods. In all cases, the problem is
  described by an effective point particle moving in a potential well with
  exponentially steep walls. Focusing on the closed-space cosmological model
  (Bianchi type IX), the mixmaster dynamics is now completely dominated by the
  quadratic Cotton tensor potential term for very small volume of the universe.
  Unlike general relativity, where the evolution towards the initial singularity
  always exhibits chaotic behavior with alternating Kasner epochs, the anisotropic
  universe in Ho\v{r}ava-Lifshitz gravity (with parameter $\lambda > 1/3$) is
  described by a particle moving in a frozen potential well with fixed (but
  arbitrary) energy $E$. Alternating Kasner epochs still provide a good description
  of the early universe for very large $E$, but the evolution appears to be
  non-ergodic. For very small $E$ there are harmonic oscillations around the fully
  isotropic model. The question of chaos remains open for intermediate energy levels.

\end{titlepage}

\vskip1cm


\section{Introduction}
\setcounter{equation}{0}

General relativity predicts the existence of space-time singularities under some
general conditions, which in a cosmological context are space-like and correspond
to the initial singularity of the universe, \cite{hawk1}, \cite{hawk2}.
Studying the asymptotic behavior of Einstein equations in the vicinity of
space-like singularities provides the way that the initial state of the universe
is reached classically. A broad framework for this purpose was developed in the
seminal work of
Belinskii, Khalatnikov and Lifshitz, \cite{bkl1}, \cite{bkl2}, \cite{bkl3},
who found that the spatial points decouple leading to dimensional reduction
of the field equations to one (time) coordinate. Then, the universe is described
as a point particle moving in an effective potential well, and, remarkably, the
dynamical equations are the same as in spatially homogeneous (but generally
non-isotropic) cosmological models, in particular Bianchi IX for having a closed
universe with spherical topology. Thus, in this context, it becomes important to
investigate the main features of the homogeneous and anisotropic cosmologies in
the small volume limit of the universe, where the matter sources are ignored.

The dynamics of Bianchi IX model in vacuum (also known as mixmaster universe) has
been thoroughly analyzed in the literature by Belinskii, Khalatnikov and Lifshitz,
\cite{bkl1}, \cite{bkl2}, \cite{bkl3}, and independently by Misner who used
Hamiltonian methods, \cite{misner1}, \cite{misner2}, \cite{misner3}, \cite{misner4},
\cite{misner5} (but see also the textbook \cite{misner6} and the monograph
\cite{ryan}). The results can be summarized in a nut-shell by saying that the
evolution consists of alternating Kasner epochs, acting as oscillations that
permute the principal axes of the spherical spatial slices, as the universe
is approaching the initial singularity. A good picture of the dynamics close to the
singularity is then provided by a billiard motion in a finite region of Lobachevsky
plane, which turns out to be chaotic. More general studies of the chaotic behavior
of the mixmaster universe have been carried out in the literature over the years,
\cite{barrow1}, \cite{barrow2}, \cite{rugh}, \cite{levin}, and
the whole subject is now well established and understood for general relativity in
four space-time dimensions. Several generalizations have also been considered
in detail, in particular in the context of higher dimensional Kaluza-Klein
theories of gravity, \cite{marc1}, \cite{marc2}, where the billiard picture
appears to be universal and the criteria for the appearance of chaos can be
formulated in Lie algebraic terms that depend on the dimensionality of
space-time, \cite{damour1}, \cite{damour2}, \cite{damour3} (but see also
\cite{damour4} for a comprehensive review of these matters and references
therein).
In parallel, there have also been studies of the same problem in higher
curvature generalizations of gravity, in particular in four dimensions, by
adding $R^2$ (and possibly other) curvature terms to the gravitational action,
where the chaotic behavior was found to be  absent, \cite{barrow3}, \cite{barrow4},
\cite{cotsakis}. Thus, the subject is quite rich and interesting in all
generally covariant effective gravitational theories, including those that
arise from string theory.

Recently, there has been a rather odd proposal in the literature to replace
the relativistic theory of Einstein gravity by a non-relativistic field
theory of Lifshitz type that is only applicable to the ultra-violet regime,
\cite{horava1}, \cite{horava2}.
The resulting theory became known as Ho\v{r}ava-Lifshitz gravity and it is
by construction a higher derivative modification of ordinary general
relativity with anisotropic scaling in the space and time coordinates.
As such, its field equations contain second derivatives in time and higher
derivatives in space coordinates (actually up to six in four space-time dimensions
where the present work will focus). This proposal aims to provide a
renormalizable theory of quantum gravity at short distances that flows to
ordinary general relativity in the infra-red domain of large distances.
It is, however, quite different in nature from the (more conventional) higher
derivative generalizations of Einstein gravity that have been considered so
far, which remain fully covariant, whereas here the modification by higher
curvature terms affects only the spatial dimensions.

It should be said straight from the beginning that there are no general theorems
for the existence of singularities in Ho\v{r}ava-Lifshitz gravity, and under
which conditions these may be valid, in particular for the existence of an initial
space-like singularity in a cosmological context\footnote{The only case that
has already been extensively studied in the literature is the analogue of
Friedmann universe,  which is an isotropic Bianchi IX model with suitable matter
sources, and was found to exhibit a bounce - rather than a singularity - in
the past under some technical but rather general conditions on the couplings
of the theory, \cite{calca}, \cite{elias}, \cite{rob}.
Such a bounce, however, might only be attributed to the particular
model, since the addition of shearing components, due to anisotropies, could
circumvent it and render it unstable. We will discuss more extensively this
point later, in section 4,
offering also some simple anisotropic models that do not exhibit a bounce.}.
Furthermore, no analysis has been made so far on how these singularities,
if present, will be approached
asymptotically at very early times (one may phrase it by simply saying that
the analysis "BKL for HL" is still lacking).

Remarkably, it can be seen that all
spatially homogeneous (but generally anisotropic) cosmological space-times,
including, in particular, the Bianchi IX model, provide consistent
mini-superspace truncations of the field equations in Ho\v{r}ava-Lifshitz
gravity, as in general relativity. In this paper we begin studying these models
in detail, first because they are interesting in their own right, as they
can provide a basis for comparing the two different theories of gravity in
the classical and (hopefully in the future) in their quantum regime, and,
second because they can also play a role in understanding the evolution of the
universe close to the initial singularity, as in general relativity. In all
cases, if the spatial points decouple close to the singularity, which is a
reasonable expectation in general, the closed universe will be effectively
described by mixmaster dynamics, viewed as point particle moving in a
potential well whose structure depends on the theory. The framework that will
be adopted throughout this study is that of Hamiltonian dynamics, since
the action of Ho\v{r}ava-Lifshitz gravity is only defined through the
$3+1$ ADM decomposition of the space-time metric.

The rest of this paper is organized as follows: Section 2 provides a
brief overview of Ho\v{r}ava-Lifshitz gravity in $3+1$ dimensions and introduces
the necessary notions in self-contained way. Section 3 explains why the
homogeneous cosmologies are consistent models in Einstein gravity as well as in
Ho\v{r}ava-Lifshitz gravity and describes mixmaster dynamics as an effective
particle model moving in a potential well that is applicable to both theories.
The effective potentials are derived in each case separately. Section 4
investigates the evolution of the universe close to the initial singularity,
where the problem simplifies considerably, but it still exhibits rich structure.
Unlike general relativity, where the potential vanishes for generic values of
the anisotropy parameters and the evolution towards the initial singularity
proceeds in an oscillatory fashion with alternating Kasner epochs, the
universe in Ho\v{r}ava-Lifshitz gravity (with parameter $\lambda > 1/3$) is
described by a particle moving in a frozen potential well with prescribed (but
arbitrary) energy. The question of chaos in the corresponding motion is briefly
addressed. Finally, section 5 contains our conclusions and discusses some
open questions and directions for further research. Two appendices are also included
at the end. The first contains useful formulas for the Bianchi IX model of
three-geometries, collecting, in particular, the expressions for its Ricci
curvature and Cotton tensors. The second contains a derivation of the bounce law
from the exponentially steep walls of the potential well that will be used in
mixmaster dynamics.

\section{Ho\v{r}ava-Lifshitz gravity}
\setcounter{equation}{0}

General relativity as well as Ho\v{r}ava-Lifshitz gravity are formulated in a similar
fashion using the ADM decomposition of the four-dimensional metric in space-time $M_4$,
which is assumed to be topologically $R \times \Sigma_3$,
\be
ds^2 = -N^2 dt^2 + \gamma_{ij} (dx^i + N^i dt)(dx^j + N^j dt) ~.
\ee
The three-dimensional slices $\Sigma_3$ have metric $\gamma_{ij}$ and extrinsic
curvature tensor
\be
K_{ij} = {1 \over 2N} \left({\partial \gamma_{ij} \over \partial t} - \nabla_i N_j -
\nabla_j N_i \right) .
\ee
The space of all three-dimensional metrics $\gamma_{ij}$, which is known as
Wheeler-DeWitt superspace, is very important in this study. It is endowed with a
metric, often called DeWitt metric, \cite{dewitt}, which is taken here to depend on a
parameter $\lambda$ in general. The metric in superspace, and its inverse, are defined
as usual,
\be
{\cal G}^{ijkl} = {1 \over 2} \left(\gamma^{ik} \gamma^{jl} + \gamma^{il}
\gamma^{jk} \right) - \lambda \gamma^{ij} \gamma^{kl} ~,
\ee
and
\be
{\cal G}_{ijkl} = {1 \over 2} \left(\gamma_{ik} \gamma_{jl} + \gamma_{il}
\gamma_{jk} \right) - {\lambda \over 3 \lambda - 1} \gamma_{ij} \gamma_{kl} ~,
\ee
so that
\be
{\cal G}_{ijmn} {\cal G}^{mnkl} = {1 \over 2} \left(\delta_i^k \delta_j^l +
\delta_i^l \delta_j^k \right) ~.
\ee
The DeWitt metric is positive definite for $\lambda < 1/3$ and indefinite for
$\lambda > 1/3$, which includes, in particular, the special value $\lambda = 1$
applicable to Einstein gravity.

The gravitational theories that will be considered in the sequel, using the ADM
formalism (see, for instance, the textbook \cite{misner6}), admit a
four-dimensional action
\be
S = S_{\rm K} - S_{\rm V} ~,
\ee
where $S_{\rm K}$ is the kinetic part of the action with universal form
\be
S_{\rm K} = {2 \over \kappa^2} \int dt d^3 x \sqrt{{\rm det} \gamma} ~ N ~
K_{ij} {\cal G}^{ijkl} K_{kl} ~.
\ee
The potential part of the action, $S_{\rm V}$, is given by
\be
S_{\rm V} = \int dt d^3 x \sqrt{{\rm det} \gamma} ~ N ~
{\cal V} ~,
\ee
where ${\cal V}$ is chosen according to the theory.

Ordinary general relativity corresponds to $\lambda = 1$ and has potential term
\be
{\cal V}_{\rm GR} = - {2 \over \kappa^2} (R - 2 \Lambda)
\ee
that involves second derivatives in the space coordinates. Here, $\Lambda$ is the
cosmological constant in $M_4$, $R$ is the Ricci scalar curvature of the
three-dimensional metric $\gamma_{ij}$ and $\kappa^2 = 32 \pi G$ is expressed
in terms of Newton's
constant in four space-time dimensions. On the other hand, Ho\v{r}ava-Lifshitz
gravity has a potential that involves higher order terms, thus breaking
relativistic invariance of the four-dimensional theory, \cite{horava1},
\cite{horava2}. These terms have a
specific form, composed of several higher order (quadratic) curvature corrections,
which are designed to smooth out the ultra-violet behavior of gravity. Also, the
parameter $\lambda$ is left undetermined in this context and may run with the
energy scale in quantum theory.

A particularly simple choice of ${\cal V}$ in Ho\v{r}ava-Lifshitz gravity, though
by no means unique, corresponds to the so called "detailed balance" condition,
meaning
\be
{\cal V}_{\rm HL} = {\kappa^2 \over 2} E^{ij} {\cal G}_{ijkl} E^{kl} ~,
\ee
so that ${\cal V}$ is derived from a superpotential ${\cal W}$ in the sense
\be
E^{ij} = {1 \over 2 \sqrt{{\rm det} \gamma}} {\delta  {\cal W} \over \delta
\gamma_{ij}} ~.
\ee
In $3+1$ dimensions that will be considered here, the superpotential is taken to
be the Euclidean action of three-dimensional topological gravity on $\Sigma_3$ with
cosmological constant $\Lambda_{\rm w}$ (other than $\Lambda$),
\be
{\cal W} = {1 \over w^2}{\cal W}_{\rm CS} + \mu {\cal W}_{\rm EH} ~,
\ee
where the first term refers to the gravitational Chern-Simons action, \cite{cs},
\be
{\cal W}_{\rm CS} = \int_{\Sigma_3} d^3 x \sqrt{{\rm det} \gamma} ~
\epsilon^{ijk} \Gamma_{im}^l \left(\partial_j \Gamma_{kl}^m + {2 \over 3} \Gamma_{jn}^m
\Gamma_{kl}^n \right) ,
\ee
with $\epsilon^{123} = 1$, and the second term is the corresponding three-dimensional
Einstein-Hilbert action
\be
{\cal W}_{\rm EH} = \int_{\Sigma_3} d^3 x \sqrt{{\rm det} \gamma} ~ (R - 2
\Lambda_{\rm w}) ~.
\ee

Thus, ${\cal V}_{\rm HL}$ with "detailed balance" follows by computing $E^{ij}$
varying the superpotential functional ${\cal W}$ with respect to the metric
$\gamma_{ij}$. The result reads
\be
E^{ij} = {1 \over w^2} C^{ij} - {\mu \over 2} G^{ij} ~,
\ee
where $C_{ij}$ is the Cotton tensor of the metric $\gamma_{ij}$ defined as follows,
\be
C^{ij} = {1 \over 2 \sqrt{{\rm det} \gamma}} {\delta {\cal W}_{\rm CS} \over \delta
\gamma_{ij}} =
{1 \over \sqrt{{\rm det} \gamma}} \epsilon^{ikl} \nabla_k \left({R^j}_l
- {1 \over 4} {\delta^j}_l R \right) .
\ee
This is a symmetric and traceless tensor that vanishes if and only if the
three-dimensional metric is conformally flat. The second term is the familiar
Einstein tensor on $\Sigma_3$ with three-dimensional cosmological constant
$\Lambda_{\rm w}$,
\be
G^{ij} = - {1 \over \sqrt{{\rm det} \gamma}} {\delta {\cal W}_{\rm EH} \over \delta
\gamma_{ij}} =
R^{ij} - {1 \over 2} R \gamma^{ij} + \Lambda_{\rm w} \gamma^{ij} ~.
\ee

Putting all together, the potential terms of Ho\v{r}ava-Lifshitz gravity in $3+1$
dimensions satisfying the "detailed balance" condition are
\be
{\cal V}_{\rm HL} = \alpha C_{ij} C^{ij} + \beta C_{ij} R^{ij}
+ \gamma R_{ij} R^{ij} + \delta R^2 + \epsilon R + \zeta
\label{vhl}
\ee
with coefficients
\ba
& & \alpha = {\kappa^2 \over 2 w^4} ~, ~~~~~ \beta = - {\mu \kappa^2 \over 2 w^2} ~,
~~~~~ \gamma = {\mu^2 \kappa^2 \over 8} ~, ~~~~~ \delta = - {\mu^2 \kappa^2
(4 \lambda - 1) \over 32(3 \lambda -1)} ~, \nonumber\\
& & \epsilon = {\mu^2 \kappa^2 \Lambda_{\rm w} \over 8(3 \lambda -1)}
~, ~~~~~ \zeta = - {3 \mu^2 \kappa^2 \Lambda_{\rm w}^2 \over 8(3\lambda -1)}~.
\label{detbala}
\ea
The last two terms are identical to the potential ${\cal V}_{\rm GR}$ of general
relativity, with the appropriate identifications of the coefficients $\epsilon$ and
$\zeta$, whereas the remaining ones are higher order (quadratic) curvature corrections
that apparently are suppressed in the infra-red limit of the theory.

More general choices of ${\cal V}_{\rm HL}$, other than "detailed balance", are also
admissible and correspond to the sum \eqn{vhl} with arbitrary coefficients; they are
only subject to the restriction that general relativity will emerge in the infra-red
regime. The analysis that will be performed in the sequel applies equally well to all
such general choices of potential in Ho\v{r}ava-Lifshitz gravity with or without
"detailed balance".

Finally, it is important to note that the Ho\v{r}ava-Lifshitz theory of gravity is not
invariant under general coordinate transformations in space-time; this should be contrasted
with other higher order theories of gravity that remain relativistic. Since $M_4$ is
topologically $R \times \Sigma_3$, it is only appropriate here to consider invariance of the
action under the restricted class of foliation preserving diffeomorphisms,
\be
\tilde{t} = \tilde{t} (t) ~, ~~~~~~ \tilde{x}^i = \tilde{x}^i (t, x) ~.
\ee
Thus, the lapse function $N$ associated to the freedom of time reparametrization is
restricted to be a function of $t$ alone, whereas the shift vector $N_i$ associated with 
diffeomorphisms of $\Sigma_3$ can depend on all space-time coordinates.

\section{Mixmaster dynamics}
\setcounter{equation}{0}

The mixmaster universe arises as mini-superspace model of gravity assuming that the
the three-dimensional slices $\Sigma_3$ are homogeneous geometries with the topology
of $S^3$ and isometry group $SU(2)$. Thus, by employing the Bianchi IX ansatz, as
explained in Appendix A, the four-dimensional metric is given in diagonal
form\footnote{More general non-diagonal metrics of the form
$ds^2 = -N^2 (t) dt^2 + \gamma_{ij} (t) \sigma_i \sigma_j$ are also legitimate for
investigation, but they will not be considered at all in this paper.}
 by
\be
ds^2 = -N^2(t) dt^2 + \gamma_1 (t) \sigma_1^2 + \gamma_2 (t) \sigma_2^2 +
\gamma_3 (t) \sigma_3^2 ~,
\ee
using the invariant 1-forms $\sigma_i$ of $SU(2)$. The metric coefficients
are all taken to depend only on the time coordinate $t$. This class of metrics provides
consistent reduction of vacuum Einstein equations to an autonomous system of ordinary
non-linear differential equations for $\gamma_i (t)$ that has been studied extensively
in the literature in the past fourty years. It provides a simple model of homogeneous,
but generally anisotropic, universe, which proves valuable for studying the chaotic
behavior of general relativity close to the initial singularity. The same ansatz also
works consistently for the Ho\v{r}ava-Lifshitz gravity with or without "detailed
balance" and gives rise to another - though more complicated - system of ordinary
non-linear differential equations for the coefficients $\gamma_i (t)$.

The purpose of this section is to describe in detail the mini-superspace reduction of
the field equations and transform them into an effective point particle model using
Hamiltonian methods, as outlined by Misner, \cite{misner1}, \cite{misner2},
\cite{misner3}, \cite{misner4}, \cite{misner5} (but see also the textbook
\cite{misner6} and the monograph \cite{ryan}). Although our discussion is entirely
confined to the Bianchi IX
case, it should be noted here that all homogeneous spaces arising in the Bianchi
classification of model three-geometries provide consistent reduction of general
relativity as well as Ho\v{r}ava-Lifshitz gravity. The details for all other homogeneous
cosmologies in Ho\v{r}ava-Lifshitz gravity will not be included here.

\subsection{Effective particle model}

Hamiltonian methods for homogeneous cosmologies are most appropriate to use in the ADM
decomposition of space-time and they lead naturally to an effective point particle
model with appropriately chosen external potential. The method is applicable to
both general relativity and Ho\v{r}ava-Lifshitz gravity because the lapse function
$N$ is taken to depend only on $t$ in such cases.

Recall that the canonical momenta conjugate to $\gamma_{ij}$ are simply given by
\be
\pi^{ij} = {\delta S \over \delta (\partial \gamma_{ij} / \partial t)} =
{2 \over \kappa^2} \sqrt{{\rm det} \gamma} ~ {\cal G}^{ijkl} K_{kl}
\ee
given the dependence of the general gravitational action $S = S_{\rm K} - S_{\rm V}$
upon the extrinsic curvature, whereas the momenta conjugate to $N$ and $N_i$ vanish.
Then, the Hamiltonian form of the action is
\be
S = \int dt d^3 x \left(\pi^{ij} {\partial \gamma_{ij} \over
\partial t} - N {\cal H} - N_i {\cal H}^i \right) ,
\ee
where
\be
{\cal H} = {\kappa^2 \over 2 \sqrt{{\rm det} \gamma}} \pi^{ij}
{\cal G}_{ijkl} \pi^{kl} + \sqrt{{\rm det} \gamma} ~ {\cal V}
\ee
and
\be
{\cal H}^i = -2 \nabla_{j} \pi^{ij} ~.
\ee

In ordinary general relativity ${\cal H} = 0$ and ${\cal H}_i = 0$ are the constraints
of the theory associated to general reparametrization invariance in the $3+1$
decomposition of the metric. In this context, $N$ and $N_i$ serve as Lagrange
multipliers whose variation yields the constraints. On the other hand, the
invariance of Ho\v{r}ava-Lifshitz gravity under the restricted class of foliation
preserving diffeomorphisms leaves intact the momentum constraints ${\cal H}_i = 0$
and replaces the time constraint ${\cal H} = 0$ with the much weaker condition
\be
\int d^3 x {\cal H} = 0 ~.
\ee
For homogeneous cosmologies, ${\cal H}$ is only a function of $t$, and so is the lapse
function $N$, and, therefore, there is no difference in the Hamiltonian description
of the two theories other than the form of the potential. Also, in these cases,
the momentum constraints are satisfied identically, since $\gamma_{ij}$ and $\pi^{ij}$
are functions of $t$ only and the covariant derivative in ${\cal H}_i$ with respect to
space coordinates reduces to an ordinary derivative. Thus, one may consistently choose
$N_i = 0$ and forget altogether the momenta constraints.

Based on these observations, the Hamiltonian form of the gravitational action for
homogeneous cosmologies takes the following form
\be
S = 16 \pi^2 \int dt \left(\pi^{ij} {d g_{ij} \over d t}
- N {\cal H} \right)
\label{effact}
\ee
after performing the integration over space that accounts for the $16 \pi^2$ factor.
Here,
\be
{\cal H} = {\kappa^2 \over 2 \sqrt{{\rm det} \gamma}} \left(\pi^{ij} \pi_{ij} -
{\lambda \over 3\lambda -1} ({\pi^k}_k)^2 \right) + \sqrt{{\rm det} \gamma} ~
{\cal V}
\ee
depends only on $t$ and it is constrained to vanish by varying $S$ with respect to $N$.

At this point, one may employ the freedom of time reparametrizations $\tilde{t} =
\tilde{t} (t)$ to eliminate $N$ from the variational problem. Any choice of $N(t)$
inserted in equation \eqn{effact} leaves the action in canonical Hamiltonian form,
but the content of the gauge fixed action will be equivalent to the original one
only if it is supplemented by the constraint ${\cal H} = 0$,
which can no longer be derived from the variational principle. The most convenient
choice is
\be
N(t) = {6 \over \kappa^2} \sqrt{{\rm det} \gamma (t)} ~,
\label{lapse}
\ee
which will be adopted from now on. Thus, we arrive at an effective point particle
Hamiltonian model for all homogeneous cosmologies; yet another formulation of this
point particle model will be mentioned shortly.

Further simplification occurs by introducing the volume and shape moduli of the
three-geometry
\be
\gamma_1 = e^{2 \Omega + \beta_+ + \sqrt{3} \beta_-} ~, ~~~~~
\gamma_2 = e^{2 \Omega + \beta_+ - \sqrt{3} \beta_-} ~, ~~~~~
\gamma_3 = e^{2 \Omega -2 \beta_+} ~,
\ee
as explained in Appendix A, so that $\sqrt{{\rm det} \gamma} = \sqrt{\gamma_1
\gamma_2 \gamma_3} = {\rm exp}(3 \Omega)$. Then, it is appropriate to parametrize
the components of the momenta matrix ${\pi^i}_j$, which is diagonal as is the
metric matrix, as follows,
\be
p_{\Omega} = 2 {\pi^k}_k ~, ~~~~~ {p^i}_j = {\pi^i}_j - {1 \over 3} {\delta^i}_j
{\pi^k}_k ~,
\ee
and set
\be
{p^1}_1 = {1 \over 6} (p_+ + \sqrt{3} ~ p_-) ~, ~~~~~
{p^2}_2 = {1 \over 6} (p_+ - \sqrt{3} ~ p_-) ~, ~~~~~
{p^3}_3 = -{1 \over 3} p_+ ~.
\ee
It turns out that $p_{\Omega}$ is the conjugate momentum to the volume moduli
$\Omega$ and $p_{\pm}$ are the conjugate momenta to the shape moduli $\beta_{\pm}$.
In terms of these variables, the gauge fixed action $S$ becomes
\be
S = 16 \pi^2 \int dt \left(p_{\Omega} {d \Omega \over dt} + p_+ {d \beta_+ \over dt}
+ p_- {d \beta_- \over dt} - H \right) ,
\label{final1}
\ee
where $t$ denotes now the specific time coordinate following the choice \eqn{lapse}
and
\be
H = {1 \over 2} \left(p_+^2 + p_-^2 - {1 \over 2(3\lambda -1)} p_{\Omega}^2 \right)
+ V (\beta_+, \beta_-, \Omega) = 0 ~.
\label{final2}
\ee
With the given choice of lapse, the effective potential of the model is
\be
V = {6 \over \kappa^2} e^{6 \Omega} {\cal V} ~.
\ee
This is the final form of the action that will be used to study the mixmaster
universe in general relativity and Ho\v{r}ava-Lifshitz gravity and compare the
results for the two theories. Note that in writing $H$ above we left undetermined
the parameter $\lambda$ of the superspace metric in order to apply it to all cases
of present interest.

Finally, as promised, we mention for completeness that there is an alternative
formulation of the same effective point particle model based on the reduced ADM
action. In this approach, which has been mostly used by Misner (and many others) in
general relativity, one eliminates $\Omega$ by solving the Hamiltonian constraint
for $p_{\Omega}$ and sets $t = \Omega$, choosing also $N$ appropriately.
The procedure is similar to going from the quadratic form of the action of a
relativistic particle to its square-root form. Then, the reduced variational problem
corresponds to a point particle moving in two dimensions (provided by $\beta_+$ and
$\beta_-$ alone) under the influence of a time dependent potential well. The
reduced ADM formulation is only adequate for $\lambda > 1/3$, but it will not be used
here in any case.

Next, we will specialize the discussion to general relativity and Ho\v{r}ava-Lifshitz
gravity and derive, in each case separately, the effective potential of the mixmaster
universe that enters in the canonical Hamiltonian variational principle \eqn{final1},
\eqn{final2}. Hamilton's equations follow by varying $\Omega$, $\beta_{\pm}$,
$p_{\Omega}$ and $p_{\pm}$ and yield
\be
{d^2 \beta_{\pm} \over dt^2} = -{\partial V \over \partial \beta_{\pm}}  ~ ,
~~~~~~~
{d^2 \Omega \over dt^2} = {1 \over 2(3\lambda -1)} {\partial V \over \partial \Omega} ~.
\ee
We will not write down explicitly the resulting equations of motion - they cannot
be solved in any case - but focus only on the qualitative features of mixmaster
dynamics, following from the potential, which make it a valuable tool for exploring
the behavior of the universe as it approaches the initial singularity\footnote{In
quantum cosmology one implements the Hamiltonian constraint by postulating the
Wheeler-DeWitt equation $\hat{H} \Psi = 0$ for the "wave-function" of the universe
with the appropriate factor ordering prescription. Some aspects of the problem
have already been studied in the literature for the mixmaster model of ordinary
quantum gravity, \cite{quantum1}, \cite{quantum2}, \cite{quantum3}, and similar
considerations can also be applied to the case of
Ho\v{r}ava-Lifshitz canonical quantum gravity. We will postpone any further
discussion of the quantum aspects of early time cosmology to future publication.}.

\subsection{General relativity}

First, we consider the mixmaster dynamics in general relativity and present the
form of the effective particle potential for comparison later with Ho\v{r}ava-Lifshitz
gravity. It follows by expressing the three-dimensional Ricci scalar $R$, given
in Appendix A, in terms of $\Omega$ and $\beta_{\pm}$ and reads, \cite{misner1},
\cite{misner2}, \cite{misner3}, \cite{misner4}, \cite{misner5},
\be
V_{\rm GR} = {6 \over \kappa^4} e^{4 \Omega} \Big[2 e^{2 \beta_+}
\left({\rm cosh}(2\sqrt{3} \beta_-) - 1 \right) -4 e^{- \beta_+}
{\rm cosh}(\sqrt{3} \beta_-) + e^{-4 \beta_+} \Big] +
{24 \Lambda \over \kappa^4} e^{6 \Omega} ~.
\ee

The equations of motion that follow from the variational principle (setting also
$\lambda = 1$) cannot be solved exactly. However, they have been studied extensively
in the literature for many years and found to exhibit some very interesting
qualitative features close to the initial singularity, where the volume of the
universe vanishes as $\Omega \rightarrow - \infty$. These features will be discussed
in some detail later. It will also be helpful in this context to have a good
qualitative picture of the potential well.

The effective potential of mixmaster dynamics corresponds to a well shown in Fig.1
for fixed $\Omega$. It has three canyon lines located at $\beta_- = 0$ and
$\beta_- = \pm \sqrt{3} \beta_+$, where any two $\gamma_i$'s become equal. The
potential is bounded from below and exhibits discrete $Z_3$ symmetry by permuting
the principal axes of rotation of $S^3$. Thus, it has the shape of an equilateral
triangle in the anisotropy space $(\beta_+ , ~ \beta_-)$ and exponentially steep
walls far away from the origin. Very close to the origin the well is round, as
can be seen by expanding $V_{\rm GR}$ up to quadratic order in $\beta_{\pm}$.

\begin{figure}[h]\centering
\epsfxsize=10cm\epsfbox{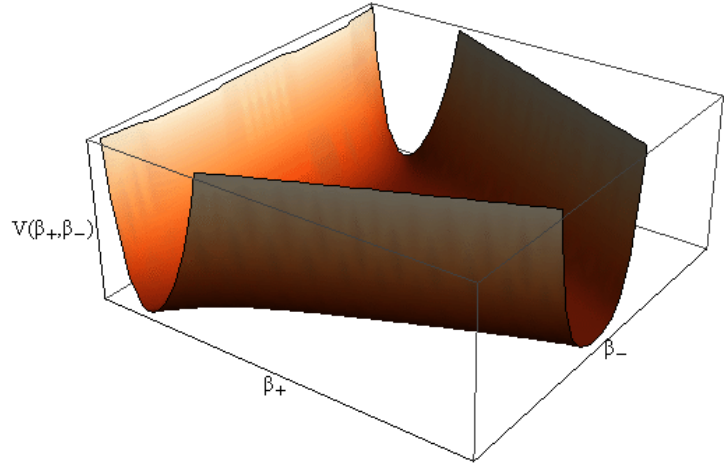}
\caption{The potential well of mixmaster dynamics and its three canyons}
\end{figure}

Another useful representation of the effective potential is shown in Fig.2 below by
drawing the equipotential curves.

\begin{figure}[h]\centering
\epsfxsize=8cm\epsfbox{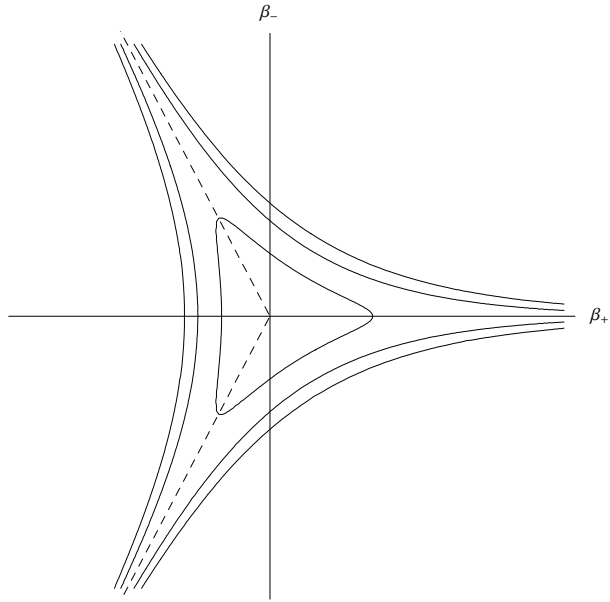}
\caption{Equipotential lines of the effective potential $V_{\rm GR} (\beta_+ ,
\beta_-)$}
\end{figure}

\noindent
As can be seen, they extend symmetrically between the
canyon lines $\beta_- = 0$ and $\beta_- = \pm \sqrt{3} \beta_+$, which correspond
to a partially anisotropic universe with axial symmetry (known as Taub space-time,
\cite{taub}).
The fully isotropic case corresponds to the origin $\beta_+ = 0 = \beta_-$, where
the potential attains its minimum.

The asymptotic form of the potential for very large values of anisotropy is
independent of the cosmological constant $\Lambda$ and looks like
\be
V_{\rm GR} \simeq {6 \over \kappa^4} e^{4 \Omega} e^{-4 \beta_+} ~, ~~~~~~~~
{\rm as} ~~~ \beta_+ \rightarrow - \infty
\label{asymin}
\ee
and
\be
V_{\rm GR} \simeq {72 \over \kappa^4} e^{4 \Omega} \beta_-^2 e^{2 \beta_+} ~,
~~~~~~~~ {\rm as} ~~~ \beta_+ \rightarrow + \infty ~, ~~~ |\beta_-| << 1 ~.
\ee
Then, the asymptotic structure of the potential is completely characterized by
combining these two relations with the triangular symmetry of the model.

The effective point particle can only escape to infinity along the canyon lines
where the potential has the shape shown in Fig.3, keeping $\Omega$ fixed. The
smallest deviation from axial symmetry will turn the particle against the
infinitely steep walls.

\begin{figure}[h]\centering
\epsfxsize=10cm\epsfbox{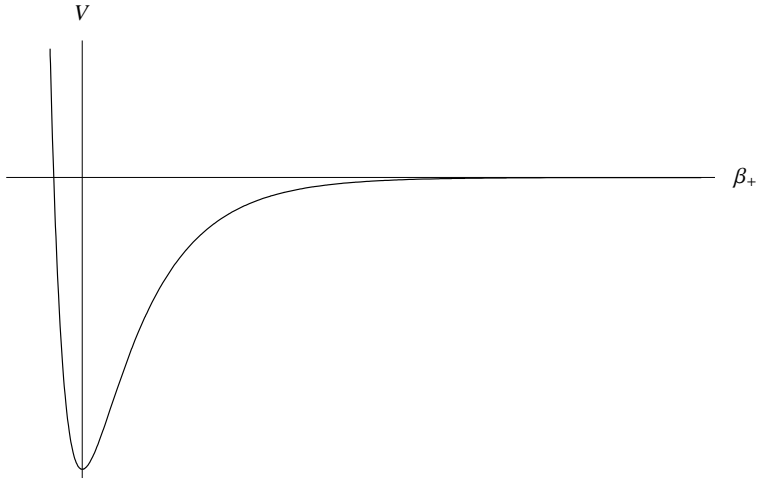}
\caption{The form of the potential $V_{\rm GR} (\beta_+ , \beta_-)$ along the canyon
line $\beta_- = 0$}
\end{figure}

\subsection{Ho\v{r}ava-Lifshitz gravity}

Applying the same framework to Ho\v{r}ava-Lifshitz gravity, the effective point particle
potential is described by
\be
V_{\rm HL} = {6 \over \kappa^2} e^{6 \Omega} \left(\alpha C_{ij} C^{ij} + \beta
C_{ij} R^{ij} + \gamma R_{ij} R^{ij} + \delta  R^2 + \epsilon R + \zeta \right) ~.
\label{genhl}
\ee
The coefficients are left arbitrary so that the discussion can be made as general
possible without necessarily imposing the "detailed balance" condition. The only
restriction we put here is $\alpha > 0$ for well-definiteness, so that $V_{\rm HL}
(\beta_+ , \beta_-)$ stays bounded from below.

Using the explicit expressions for $C_{ij}$, $R_{ij}$ and $R$ found in Appendix A and
rewriting them in terms of the variables $\Omega$ and $\beta_{\pm}$, we obtain the
following results for the individual terms entering in the effective potential of
Bianchi IX cosmology:
\ba
C_{ij}C^{ij} & = & {1 \over 2} e^{-6 \Omega} \Big[2 e^{6 \beta_+}
\left(3 {\rm cosh}(6\sqrt{3} \beta_-)
- 3 {\rm cosh}(4\sqrt{3} \beta_-) + {\rm cosh}(2\sqrt{3} \beta_-) - 1\right) -
\nonumber\\
& & 2 e^{3 \beta_+} \left(3 {\rm cosh}(5\sqrt{3} \beta_-) - {\rm cosh}(3\sqrt{3}
\beta_-) - 2 {\rm cosh}(\sqrt{3} \beta_-) \right) + \nonumber\\
& & 2 \left({\rm cosh}(4\sqrt{3} \beta_-) + 2 {\rm cosh}(2\sqrt{3} \beta_-) -
3 \right) - \nonumber\\
& & 4 e^{-3 \beta_+} \left({\rm cosh}(3\sqrt{3} \beta_-) - {\rm cosh}(\sqrt{3}
\beta_-) \right) + \nonumber\\
& & e^{-6 \beta_+} \left(2 {\rm cosh}(2\sqrt{3} \beta_-) + 1\right) - \nonumber\\
& & 6 e^{-9 \beta_+} {\rm cosh}(\sqrt{3} \beta_-) + 3 e^{-12 \beta_+} \Big] ~,
\ea
\ba
C_{ij}R^{ij} & = & - e^{-5 \Omega} \Big[2 e^{5 \beta_+} \left({\rm cosh}(5\sqrt{3}
\beta_-) - {\rm cosh}(3 \sqrt{3} \beta_-) \right) \nonumber\\
& & - 2 e^{2 \beta_+} \left({\rm cosh}(4\sqrt{3} \beta_-)
- {\rm cosh}(2 \sqrt{3} \beta_-) \right) + e^{-4 \beta_+} \nonumber\\
& & - 2 e^{-7 \beta_+} {\rm cosh}(\sqrt{3} \beta_-) + e^{-10 \beta_+} \Big] ~,
\ea
\ba
R_{ij}R^{ij} & = & {1 \over 4} e^{-4 \Omega} \Big[2 e^{4 \beta_+}
\left(3 {\rm cosh}(4\sqrt{3} \beta_-)
- 4 {\rm cosh}(2 \sqrt{3} \beta_-) + 1 \right) - \nonumber\\
& & 8 e^{\beta_+} \left({\rm cosh}(3\sqrt{3} \beta_-)
- {\rm cosh}(\sqrt{3} \beta_-) \right) + \nonumber\\
& & 4 e^{-2 \beta_+} \left({\rm cosh}(2\sqrt{3} \beta_-) + 1 \right) - \nonumber\\
& & 8 e^{-5 \beta_+} {\rm cosh}(\sqrt{3} \beta_-) + 3 e^{-8 \beta_+} \Big] ~,
\ea
\ba
R^2 & = & {1 \over 4} e^{-4 \Omega} \Big[2 e^{4 \beta_+}
\left({\rm cosh}(4\sqrt{3} \beta_-)
- 4 {\rm cosh}(2 \sqrt{3} \beta_-) + 3 \right) - \nonumber\\
& & 8 e^{\beta_+} \left({\rm cosh}(3\sqrt{3} \beta_-)
- {\rm cosh}(\sqrt{3} \beta_-) \right) + \nonumber\\
& & 4 e^{-2 \beta_+} \left(3 {\rm cosh}(2\sqrt{3} \beta_-) + 1 \right) - \nonumber\\
& & 8 e^{-5 \beta_+} {\rm cosh}(\sqrt{3} \beta_-) + e^{-8 \beta_+} \Big]
\ea
and
\be
R = - {1 \over 2} e^{-2 \Omega} \Big[2 e^{2 \beta_+}
\left({\rm cosh}(2\sqrt{3} \beta_-) - 1
\right) -4 e^{- \beta_+} {\rm cosh}(\sqrt{3} \beta_-) + e^{-4 \beta_+} \Big] ~.
\ee

The potential $V_{\rm HL} (\beta_+ , \beta_-)$ also has the shape of an equilateral
triangle with exponentially steep walls when $\alpha >0$. Fig.1 and Fig.2 still provide
a good qualitative picture of it far from the origin in $(\beta_+ , \beta_-)$ parameter
space. Only the bottom area close to the origin has slightly different
shape that depends on the relative coefficients of the individual terms of the potential.
The equations of motion that follow from it provide the analogue of mixmaster dynamics in
Ho\v{r}ava-Lifshitz gravity. We will not attempt to solve them here but rather confine
ourselves to study some qualitative features that make the model useful for early time
cosmology, as in general relativity.

For $\alpha >0$, which will be assumed from now on, the asymptotic form of
the potential $V_{\rm HL}$ for very large
values of anisotropy is dominated completely by the quadratic Cotton tensor
term, which happens to contain the steepest walls of all terms, and one has
\be
V_{\rm HL} \simeq {9 \over \kappa^2} \alpha e^{-12 \beta_+} ~, ~~~~~~~~
{\rm as} ~~~ \beta_+ \rightarrow - \infty ~.
\ee
Likewise, we have
\be
V_{\rm HL} \simeq {576 \over \kappa^2} \alpha \beta_-^2 e^{6 \beta_+} ~,
~~~~~~~~ {\rm as} ~~~ \beta_+ \rightarrow + \infty ~, ~~~ |\beta_-| << 1 ~.
\ee
Thus, unlike general relativity, we note that the asymptotic form of the potential
is independent of $\Omega$.

As before, the effective point particle can only escape to infinity along the canyon
lines $\beta_- = 0$ and $\beta_- = \pm \sqrt{3} \beta_+$ which arise for general
choices of $V_{\rm HL}$.

\begin{figure}[h]\centering
\epsfxsize=10cm\epsfbox{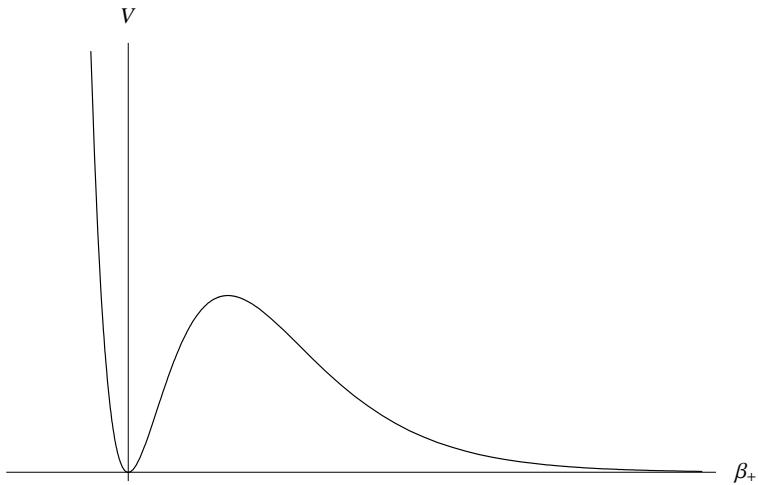}
\caption{The form of the potential $V_{\rm Cotton} (\beta_+ , \beta_-)$ along the canyon
line $\beta_- = 0$}
\end{figure}

\noindent
Fig.4 shows only the plot of the quadratic Cotton term of the potential, $V_{\rm Cotton}$,
along one of these lines, say $\beta_- = 0$; it exhibits a local maximum at $\beta_+ =
({\rm log} 2)/3$, where $\gamma_1 = \gamma_2 = 2 \gamma_3$, and similar local maxima
show up along the other two lines obtained by permuting $\gamma_i$. As will be seen
later $V_{\rm Cotton}$ is the only relevant term of $V_{\rm HL}$ in early time
cosmology.

\section{Approach to the initial singularity}
\setcounter{equation}{0}

In this section we examine the dynamical behavior of the universe close to the
initial singularity, where $\Omega \rightarrow -\infty$, using the Bianchi IX
model in vacuum. In this context, it is important to have anisotropic models
with general parameters $\beta_{\pm}$, since, otherwise, the universe will not
be able close up to $S^3$ without radiation or matter sources.
First, we will make some general - though crude - remarks about the existence
of the initial singularity and then study the problem in question for the two
different theories of gravitation.

\subsection{General considerations}

The isotropic Bianchi IX case in vacuum corresponds to a closed Robertson-Walker
space-time
\be
ds^2 = - N^2 (t) dt^2 + e^{2 \Omega (t)} d\Omega_3^2 ~,
\ee
with $N(t) = (6/\kappa^2){\rm exp} (3\Omega)$. Note, however, that this metric
cannot remain isotropic more than instantaneously; consistency of
the dynamics also requires the presence of some shearing components
provided by space anisotropy in vacuum - beyond the pure dilation - or suitable
sources and combinations thereof.

In general relativity, this follows from the form of the potential assuming
$\beta_+ = 0 = \beta_-$ for all time. Since the potential
\be
V_{\rm GR}^{\rm isotropic} (\Omega) = -{6 \over \kappa^4} \Big[3 e^{4 \Omega}
- 4 \Lambda e^{6 \Omega} \Big]
\ee
is always negative for small volume, irrespective of $\Lambda$, it fails to
satisfy the Hamiltonian constraint (setting $\lambda = 1$),
\be
{1 \over 8} p_{\Omega}^2 = 2 \left({d \Omega \over dt} \right)^2 =
V_{\rm GR}^{\rm isotropic} (\Omega) ~.
\ee

Adding sources, in the form of perfect fluid, remedies the situation and
yields the Friedmann universe. Recall in this case that the potential
density ${\cal V}_{\rm GR}$ is modified by adding the (positive) contribution
of the energy density $\rho$ of the fluid, so that\footnote{This is one of
the Friedmann equation in standard cosmology with $p_{\Omega}$ being
the Hubble parameter. Also, to compare with the more standard form of these
equations, it is appropriate to use another time frame defined as
$dT = N(t) dt$, where the Robertson-Walker metric takes the more familiar
form $ds^2 = -dT^2 + a^2(T) d\Omega_3^2$ with $a = {\rm exp} \Omega$. Then,
the initial singularity occurs at some finite past proper time, say
$T=0$, instead of $t = -\infty$.}
\be
2 \left({d \Omega \over dt} \right)^2 = V_{\rm GR} +
\mu ~ e^{\nu ~ \Omega} ~,
\ee
where the contribution of non-relativistic matter sources corresponds to $\nu = 3$
(since $\rho \sim V^{-1}$) and that of radiation to $\nu = 2$ (since
$\rho \sim V^{-4/3}$) with
$V = {\rm exp} (3 \Omega)$. Clearly, the radiation term provides the dominant
contribution in the small volume limit and one obtains an expanding isotropic
and homogeneous universe with initial singularity at $T=0$. If deviations from
isotropy are subdominant in the small volume limit, compared to radiation, the
initial singularity will persist in the presence of sources. This argument
only applies to the anisotropy terms of the potential $V_{\rm GR}$ that depends
on the volume as
${\rm exp}(4 \Omega)$ via the curvature. The anisotropy, however, introduces
additional (positive definite) terms in the Hamiltonian constraint, namely the
kinetic energy of the anisotropy parameters $(d\beta_+ / dt)^2 + (d\beta_- / dt)^2$
if one considers the Bianchi IX model, which completely dominate the evolution
close to the initial singularity - essentially trying to "avoid" it by
oscillations - as will be discussed later. In all cases, the universe can come
to a singular state at $T=0$ satisfying the field equations.

In Ho\v{r}ava-Lifshitz gravity, the isotropic potential does not receive
contribution from the Cotton tensor, since $C_{ij}$ vanishes identically. Thus,
the potential given in general by equation \eqn{genhl}, without
necessarily assuming "detailed balance" condition, becomes
\be
V_{\rm HL}^{\rm isotropic} (\Omega) = {3 \over 2 \kappa^2} \Big[3 (\gamma +
3 \delta) e^{2 \Omega} + 6 \epsilon e^{4 \Omega} + 4 \zeta e^{6 \Omega}
\Big] ~.
\ee
The first term arises from the combined effect of $R_{ij} R^{ij}$ and $R^2$
and dominates the dynamics for small volume. The Hamiltonian constraint now
reads
\be
{1 \over 4(3 \lambda - 1)} p_{\Omega}^2 = (3 \lambda - 1)
\left({d \Omega \over dt} \right)^2 = V_{\rm HL}^{\rm isotropic} (\Omega)
\ee
and cannot be possibly fulfilled when
\be
(3 \lambda - 1) (\gamma + 3 \delta) < 0 ~.
\label{roula}
\ee
This inequality, which is certainly satisfied for in the case of "detailed balance"
condition (see the choice of coefficients \eqn{detbala}), means that the quadratic
curvature terms correspond to "dark radiation" (since they effectively have
"$\nu = 2$"), but with negative energy density. It also implies
that the universe cannot evolve isotropically in vacuum without turning on
some shearing components, as in general relativity.

Adding sources, in the form of perfect fluid, leads
to an interesting possibility when inequality \eqn{roula} is fulfilled with
$\lambda > 1/3$. In analogy with the previous analysis one obtains
\be
(3 \lambda - 1) \left({d \Omega \over dt} \right)^2 = V_{\rm HL} +
\mu ~ e^{\nu ~ \Omega}
\ee
and the dominant contribution in the small volume limit is provided by the
quadratic Ricci curvature terms and the matter sources with suitable $\nu$.
Then, isotropic evolution becomes possible, leading to a Friedmann universe
in Ho\v{r}ava-Lifshitz gravity. When $\nu > 2$, there can be a bounce in $\Omega$
that replaces the initial singularity of the universe, \cite{calca}, \cite{elias},
\cite{rob}, as can be easily seen by neglecting the contribution of
the curvature $R$ and the cosmological constant term.
This is the only case for which the energy density of "dark radiation" can
grow with respect to the regular matter energy\footnote{This condition by itself is
quite restrictive since it rules out regular radiation before the bounce.}.
Note, however, that possible deviations from isotropy
will become dominant in the small volume limit, since the quadratic Cotton
tensor term, which is independent of $\Omega$, kicks in $V_{\rm HL}$ and washes
away the effect of the previously thought relevant terms. The kinetic energy of
the anisotropy parameters also contributes on equal footing. This indicates
that the cosmological bounce is unstable against anisotropy, and, generically,
the universe can come in a singular state at $T=0$ satisfying the field equations.
The validity or not of inequality \eqn{roula} becomes irrelevant in the
presence of anisotropy. Consistency also requires $\lambda > 1/3$, otherwise the
Hamiltonian constraint cannot be possibly satisfied in the small volume limit in
the presence of anisotropy; by the same token, the universe can only remain
still in an isotropic state when $\lambda < 1/3$, and, therefore, this possibility
will not be considered further.

Although the argument above does not provide a rigorous proof for the existence
of an initial singularity in Ho\v{r}ava-Lifshitz cosmology, and under which
general conditions this may be possible,
it seems sufficient for the purposes of the present
work\footnote{Another example for having an initial singularity - rather
than a bounce - is provided by the anisotropic Bianchi I model (also
known as Kasner solution), although the reasoning is slightly different here.
This is a common solution to general relativity and Ho\v{r}ava-Lifshitz gravity
in vacuum because $\Sigma_3$ is a flat three-dimensional space and all components
of the Ricci curvature and Cotton tensor vanish.  In this case, the only
contribution to the Hamiltonian constraint (neglecting the cosmological constant
term for small volume) is provided by the kinetic energy of the
anisotropy parameters and the universe can evolve towards the initial singularity
without reaching a minimum volume. Thus, the bounce in the
Friedmann model does not appear to be generic, in particular in the presence of
anisotropy.}. Thus, in the following, we will use mixmaster cosmology to explore
the approach to the initial singularity, as in general relativity.

\subsection{General relativity}

The potential $V_{\rm GR}$ appears to vanish as one approaches the initial
singularity. This is true for generic values of the anisotropy
parameters $\beta_{\pm}$ implying Kasner behavior of the universe close to the
singularity, which is taken to occur at the beginning of cosmic time, \cite{bkl1},
\cite{bkl2}, \cite{bkl3}, \cite{misner1}, \cite{misner2}, \cite{misner3},
\cite{misner4}, \cite{misner5}. In fact, since the Ricci scalar curvature
of the homogeneous space $\Sigma_3$ vanishes in this limit, the space is effectively
flat, as in Bianchi type-I cosmology, and it is more appropriate to
use Cartesian $dx$, $dy$ and $dz$ instead of the 1-forms $\sigma_i$ of $SU(2)$.

More precisely, when the potential vanishes, all momenta are constant satisfying
$p_{\Omega}^2 = 4(p_+^2 + p_-^2)$ by the Hamiltonian constraint (with $\lambda = 1$).
Then, it is convenient to introduce the following parametrization of the constant
momenta,
\ba
& & n_1 = {1 \over 3 p_{\Omega}} \left(p_{\Omega} -2 p_+ - 2\sqrt{3} p_- \right) ~,
~~~~~
n_2 = {1 \over 3 p_{\Omega}} \left(p_{\Omega} - 2 p_+ + 2 \sqrt{3} p_- \right) ~,
~~~~~ \nonumber\\
& & n_3 = {1 \over 3 p_{\Omega}} \left(p_{\Omega} + 4p_+ \right) ~,
\label{choic}
\ea
so that
\be
n_1 + n_2 + n_3 = 1 = n_1^2 + n_2^2 + n_3^2 ~.
\label{algcon}
\ee
The remaining equations $d\beta_{\pm} / dt = p_{\pm}$ and $d \Omega /dt = - p_{\Omega}/4$
(so that $d \beta_{\pm} / d \Omega = - 4 p_{\pm} / p_{\Omega}$) can be easily solved to
yield the metric coefficients $\gamma_i (t) = T^{2n_i}$ with respect to a time frame $T$
defined by absorbing the lapse function as $dT = N(t) dt$.
Then, the metric takes the familiar Kasner form
\be
ds^2 = -dT^2 + T^{2n_1} dx^2 + T^{2n_2} dy^2 + T^{2n_3} dz^2 ~,
\ee
which describes an expanding flat universe with linearly varying volume element,
$\sqrt{{\rm det} \gamma} = T$.

Thus, the mixmaster dynamics close to the initial singularity appears to follow
the Kasner evolution with some fixed parameters $(n_1 , n_2 , n_3)$.
The Kasner universe is anisotropic as it always
contains a direction, say $z$, along which distances contract rather than expand;
this follows from the algebraic conditions \eqn{algcon}, which imply that one of the
$n_i$'s, say $n_3$, is lying in the interval $-1/3 \leq n_3 \leq 0$. We may order the
Kasner exponents as
\be
-1/3 \leq n_3 \leq 0 \leq n_2 \leq 2/3 \leq n_1 \leq 1
\ee
without loss of generality. The axially symmetric case corresponds to the choice
of parameters $n_1 = n_2 = 2/3$ and $n_3 = -1/3$ (and permutations of the axes
thereof).

This description is valid at generic points of $(\beta_+ , \beta_-)$ parameter
space, but it can break down far away from the origin when the effective point
particle experiences the
exponentially steep walls of the potential $V_{\rm GR}$. For example, when the
particle approaches one of the triangular walls arising at $\beta_+ \rightarrow
- \infty$, as it moves within the wedge $|\beta_-| < -\sqrt{3} \beta_+$, the
dominant term of the potential is proportional to ${\rm exp} [4(\Omega - \beta_+)]$,
as can be seen from the asymptotic behavior \eqn{asymin}, and will become
sufficiently large to influence the motion if $d \beta_+ / d \Omega > 1$\footnote{The
rate $d \beta_+ / d \Omega$ is positive because $\beta_+$ decreases as $\Omega$
decreases while the particle is heading towards the wall in its descent towards the
singularity.}. In the
simplest case of axially symmetric evolution towards the wall, so that $\beta_-$
stays zero along the trajectory and the Kasner exponents are $n_1 = n_2 = 2/3$ and
$n_3 = -1/3$, it is clear from the Hamiltonian constraint
$(d \beta_+ / d\Omega)^2 + (d \beta_- / d\Omega)^2 = 4$ that
$d \beta_+ / d \Omega = 2$ and the inequality for having a bounce is satisfied.
More generally, when the particle moves within the wedge $|\beta_-| < -\sqrt{3}
\beta_+$ against the wall, we have
\be
{d \beta_+ \over d \Omega} = 1 - 3n_3
\ee
and the inequality $d \beta_+ / d \Omega > 1$
is always satisfied since $n_3 < 0$. Similar considerations also
apply to all other walls by the triangular symmetry of the model.
Thus, the point particle will always bounce against the walls of the potential.

Summarizing, close to the initial singularity,
the evolution of the universe is accurately described by Kasner dynamics until
the particle hits the walls and enters into a new Kasner phase (with different
parameters, in general) after the bounce\footnote{The bounce law for the Kasner
parameters, $n_i \rightarrow n_i^{\prime}$, has been worked out in the literature
(but see also Appendix B). For a bounce against the wall at $\beta_+ \rightarrow -
\infty$, it is most easily
described as $s/3 \rightarrow 3/s$, using Misner's parametrization, \cite{misner1},
\be
n_1 = {2s (s-3) \over 3 (s^2 + 3)} ~, ~~~~~
n_2 = {2s (s+3) \over 3 (s^2 + 3)} ~, ~~~~~
n_3 = -{(s-3)(s+3) \over 3 (s^2 + 3)} ~. \nonumber
\ee
An alternative parametrization has been provided by Belinskii, Khalatnikov and
Lifshitz, \cite{bkl1}, \cite{bkl2}, \cite{bkl3}.
The axially symmetric case $n_1 = n_2 = 2/3$ and $n_3 = -1/3$ (corresponding to $s=\infty$)
is special as the particle heads to the corner of the triangular well after the bounce,
following the canyon line $\beta_- = 0$ with Kasner parameters $n_1^{\prime} = n_2^{\prime}
= 0$ and $n_3^{\prime} = 1$ (corresponding to $s=0$); this is also apparent from equations
\eqn{choic}, since $p_- = 0$ all the time and $p_+$ only reverses sign relative to
$p_{\Omega}$ after the bounce.
The fixed points of the bounce law arise for $s = \pm 3$ and correspond to Kasner parameters
$(1, 0 , 0)$ and $(0, 1, 0)$ that describe the evolution of the particle along the canyon
lines $\beta_- = \pm \sqrt{3} \beta_+$ without bounce. The bounce law from the other two
walls follows by permutation of the axes, which is equivalent to replacing $s$ by
$(s \pm 3)/(s \mp 1)$, and amounts to $(s+3)/3(s-1) \rightarrow 3(s-1)/(s+3)$ or
$(s-3)/3(s+1) \rightarrow 3(s+1)/(s-3)$ with fixed points $s = 0$ and $3$ or $s=0$ and
$-3$ respectively.}.
The bounce repeats itself again and again, in general, leading to an oscillatory behavior
of $S^3$ that alternates its three principal axes, while the universe
is approaching the initial singularity, \cite{bkl1}, \cite{bkl2}, \cite{bkl3},
\cite{misner1}, \cite{misner2}, \cite{misner3}, \cite{misner4}, \cite{misner5}.
It appears that almost all solutions obtained by
successive bounces come arbitrarily close to the corners of the well associated to
the special values of Misner parameter $s=0$ or $s= \pm 3$ (by triangular symmetry).
Then, the space-time metric comes close to $ds^2 = -dT^2 + dx^2 + dy^2 + T^2 dz^2$
(up to permutations of the axes), which is equivalent to the flat metric
$ds^2 = - d\eta^2 + d\xi^2 + dx^2 + dy^2$
by the transformation $\xi = T ~ {\rm sinh} z$ and $\eta = T ~ {\rm cosh} z$.
A new era of alternating Kasner epochs subsequently starts by permuting the axes,
and so on. Actually, this motion can be formulated as billiard
in a finite region of hyperbolic two-dimensional space, obtained by appropriate
transformation of $(\beta_+ , \beta_-)$ parameter space, and, as such, it is
chaotic (for an overview and history of the developments in the subject see, for
instance, \cite{misner5} and references therein); it also provides the origin of
the chaotic behavior exhibited by mixmaster dynamics in general, \cite{barrow1},
\cite{barrow2}, \cite{levin}.

The intuitive characterization of having chaos in early time cosmology
is that when the universe starts with a well-defined state, it will evolve towards the
singularity by going through almost all possible anisotropic stages by changing shape,
as it does in general relativity.

\subsection{Ho\v{r}ava-Lifshitz gravity}

In this case, as one approaches the initial singularity,
where $\Omega \rightarrow -\infty$, the potential does not vanish for generic values
of $\beta_{\pm}$, contrary to what happens in general relativity. Instead, it is
well approximated by
\be
V_{\rm HL} = V_{\rm Cotton} \equiv {6 \alpha \over \kappa^2} e^{6 \Omega} C_{ij}
C^{ij} ~, ~~~~~~~ {\rm as} ~~~ \Omega \rightarrow -\infty ~.
\ee
This term can be alternatively written (in more compact form) as
\be
V_{\rm Cotton} = {\alpha \over \kappa^2} \Big[\left({\partial W \over \partial
\beta_+} \right)^2 + \left({\partial W \over \partial \beta_-} \right)^2 \Big] ~,
\ee
where the corresponding superpotential is
\ba
W & = & e^{3 \beta_+} \left({\rm cosh}(3 \sqrt{3} \beta_-) -
{\rm cosh}(\sqrt{3} \beta_-) \right) - {\rm cosh}(2 \sqrt{3} \beta_-) - \nonumber\\
& & e^{- 3 \beta_+} {\rm cosh}(\sqrt{3} \beta_-) + {1 \over 2} e^{- 6 \beta_+} ~,
\ea
and it is positive definite when $\alpha > 0$. Clearly, this applies to all models of
Ho\v{r}ava-Lifshitz gravity, with or without "detailed balance".
Consistency with the Hamiltonian constraint requires that $\lambda > 1/3$, which we
assume in the following.

This is not surprising in retrospect because the quadratic Cotton
tensor term is marginal in the gravitational action and it is expected to dominate in the
ultra-violet regime of the theory. Furthermore, since $V_{\rm Cotton}$ is independent of
$\Omega$, the scale factor of the universe will evolve as a free particle with fixed (but
arbitrary) momentum $p_{\Omega}$, so that the volume of space diminishes
linearly at early times (in the appropriate time coordinate $T$).
As for $\beta_{\pm}$ that determine the shape
of the universe, they will keep changing all the time following the motion of a
particle in a frozen (time-independent) well
\be
E = {1 \over 2} \left({d \beta_+ \over dt} \right)^2 + {1 \over 2} \left({d \beta_-
\over dt} \right)^2 + V_{\rm Cotton} (\beta_+ , \beta_-)
\label{chaos}
\ee
with fixed energy level\footnote{We must require $E>0$ so that the universe is
anisotropic. Then, for small $E$, $\Omega$ diminishes slowly towards the
initial singularity, whereas for large $E$ it diminishes very fast.}
\be
E = {1 \over 4(3 \lambda -1)} p_{\Omega}^2 ~.
\ee
Thus, the dynamics appears more complicated now, compared to general relativity, and
the universe will no be - in general - in a Kasner epoch before bouncing off the
walls.

Solving this effective point particle problem is not an easy task, but one can
examine some qualitative features of the motion depending on the energy $E$.
When $E$ is very large, the potential can be approximated by zero for generic
values of $\beta_{\pm}$, since the particle looks insensitive to the small bumps at
the bottom of the well\footnote{Actually, this is an assumption which can be safely
made only for short time development of the system, based on intuition. In general,
one should also prove that these bumps, no matter how small they are compared to $E$,
do not influence much the long time development of the system after several iterations.
Such non-perturbative results, which go back to Poincar\'e and fall within the classic
theory of Kolmogorov, Arnol'd and Moser (KAM theory), \cite{kam} (but see also 
\cite{arno} and \cite{jpos} for 
comprehensive exposition), are rather difficult to establish
in classical mechanics and they will not be considered in detail in the present work.
They should be properly investigated, however, as they might change the physical
picture we are about to present after a very long time.}.
Thus, only in this case, which resembles general relativity,
the universe will be in a Kasner epoch far away from the walls (recall that the
Kasner model is a common solution to the two theories and it is insensitive to the
cosmological constant in the small volume limit). Note, however, that the
bounce law is modified\footnote{A bounce against the steady wall
occurring at $\beta_+ \rightarrow - \infty$, follows the rule of ordinary
reflections, namely the incidence and reflection angles are equal.
It is neatly described as
$s / \sqrt{3} \rightarrow \sqrt{3} / s$, using the same parametrization of the
Kasner parameters in terms of a single variable $s$ as in general relativity.
It should be compared to the bounce law $s/3 \rightarrow 3/s$ that governs
the mixmaster universe in general relativity. In the present case, the fixed
points arise at $s = \pm \sqrt{3}$ with associated Kasner exponents
$n_1 = (1 \mp \sqrt{3})/3$, $n_2 = (1 \pm \sqrt{3})/3$ and $n_3 = 1/3$,
respectively. As can be seen from the corresponding expressions in Appendix B,
$p_+ = 0$ in this case and the point particle moves parallel to the wall (the two
signs refer to the two directions of motion), until it hits the other walls.
The bounce from the other two walls follows by permutation of the axes, as usual,
which is equivalent to replacing $s$ by $(s \pm 3)/(s \mp 1)$. Then, the bounce
law yields the other fixed points $s = 3 \pm 2 \sqrt{3}$ and $s = -3 \pm 2 \sqrt{3}$
that describe a particle moving parallel to the other two walls in either direction.
Clearly, one can have a closed orbit forming equilateral triangle, which does not
seem possible in general relativity.} compared to general relativity and resumes
its standard form,
as can be seen in Appendix B. But the qualitative picture remains the same: after
the bounce the universe will enter into another Kasner epoch and so on, as in
general relativity.

The picture of a billiard is also very useful here. For very large $E$ we have
a very large region on the plane $(\beta_+, \beta_-)$ bounded by an equilateral
triangle inside which the particle moves freely with very large velocity.
By simple rescaling, one
can reformulate this problem as a particle moving freely inside an equilateral
triangle of finite size with finite energy. The motion takes place on the plane
following the standard rule of equal incidence and reflection angles. It is easy
to prove that such a billiard is not ergodic. In fact, billiards in any triangular
domain on the plane are non-ergodic when the angles of the triangle are rational
multiples of $2\pi$, since, then, all possible reflection angles along a given
path of the particle assume only a finite set of values (see, for instance,
\cite{sinai}).

This should be contrasted with the ergodic behavior of the triangular billiard
with moving walls that arises in general relativity, which can be viewed as
a fixed triangular billiard with non-standard bounce law. The equivalent
formulation of this problem in general relativity, as billiard in a compact
domain of Lobachevsky plane  with standard reflection rules, is another way of
establishing ergodicity in
that case, since the geodesic flow on surfaces with negative constant
curvature is the prime example of ergodic behavior in Hamiltonian systems,
\cite{hopf} (but see also \cite{sinai} for an overview).

In summary, for very large $E$, the motion of the effective point particle in
Ho\v{r}ava-Lifshitz gravity is not chaotic, provided, of course, that the
small bumps of the potential do not affect much the long time behavior of the
system after many bounces. It is an immediate and important consequence of the 
KAM theory that small perturbations of non-degenerate Hamiltonian systems, like 
ours, are not ergodic\footnote{At a given energy level, however, the invariant 
non-resonant (Kronecker) tori in phase space form a Cantor set, which has no 
interior points. Therefore, it is impossible to tell with finite precision whether 
a given initial position falls on an invariant torus or in a gap between such 
tori. In such cases, according to KAM theory, one can only make probabilistic 
statements for a chosen orbit to be on an invariant torus, and, hence, be stable.}. 
Thus, in this case, if the universe starts with a well-defined state, it will 
evolve towards the singularity by changing shape but without 
passing through all possible anisotropic stages.

On the other hand, for intermediate $E$, the landscape of the bottom of the
potential becomes visible to the particle and the bumps can no longer be
ignored even for short time. Thus, the evolution between the walls becomes
more complicated and (unfortunately) it cannot be described in simple terms.
Also note that the canyon lines now exhibit a small bump, as
shown in Fig.4, and, therefore, below a threshold,
\be
E_{\star} = {9 \alpha \over 16 \kappa^2} ~,
\ee
the particle cannot exit the well and head towards its corners. For $E<E_{\star}$
the motion remains bounded, provided that the anisotropies are relatively small,
and the particle oscillates around the origin (fully
isotropic model). Thus, for intermediate $E$, the dynamics of the universe close
to the singularity is very different and complex and it is not
yet clear if it remains non-ergodic. If a second integral of motion exists, the
system will be integrable, but we have not been able to find such thing.

Finally, for very small values $E$ and relatively small anistropies, the particle
moves around the minimum of the potential as a two-dimensional isotropic oscillator,
\be
E = {1 \over 2} \left({d \beta_+ \over dt} \right)^2 + {1 \over 2} \left({d \beta_-
\over dt} \right)^2 + {81 \alpha \over \kappa^2} (\beta_+^2 + \beta_-^2) ~,
\ee
which follows by expanding $V_{\rm Cotton}$ up to quadratic order around
$\beta_+ = 0 = \beta_-$. In such case the motion is integrable and chaos is obviously
absent. The universe still exhibits oscillatory behavior by changing shape in its
descent towards the initial singularity, but the evolution is not Kasner-like. The
universe passes through the isotropic model periodically with cyclic frequency
$9 \sqrt{2 \alpha} / \kappa$.

The problem is certainly very rich and should be investigated in more detail, including
numerical studies, in order to be able to make more conclusive and safe statements about
chaos in the motion for general values of $E$.  In fact, a Hamiltonian system like
\eqn{chaos} can be ergodic at certain energy levels and non-ergodic at other levels.
Here, there are also intermediate energies $E$ separating into phases the behavior
of mixmaster dynamics close to the initial singularity of the universe.

It is instructive to compare this behavior with the absence of chaos in fully covariant
higher curvature generalizations of Einstein gravity, \cite{barrow3}, \cite{barrow4},
\cite{cotsakis}, where the reasoning is different.
In the general context of $f(R)$ gravity in four space-time dimensions, there
is a well known conformal relationship between the vacuum higher derivative theory
and ordinary general relativity coupled to a scalar field
\be
\varphi = {\rm log} f^{\prime}(R)
\ee
with potential
\be
V(\varphi) = {1 \over 2 {f^{\prime}}^2} (R f^{\prime} - f) ~,
\ee
where prime denotes the derivative with respect to the four-dimensional scalar
Ricci curvature. In the simplest case, the Einstein-Hilbert Lagrangian is replaced by
$f(R) = R + \alpha R^2$, but $f(R)$ can also assume more general forms. The mixmaster
universe provides a consistent mini-superspace model of $f(R)$ gravity, which
is still described by a point particle that bounces off the walls of a triangular
potential in the small volume limit. There is also an analogue of the Kasner solution
in general relativity coupled to a scalar field, which is appropriate to use in this
case. However, the effect of the scalar field $\varphi$
is to slow down the speed of the point particle relative to the moving walls and the
particle will bounce back only if it moves not too oblique relative to the walls;
it should be contrasted to general relativity without scalar field, where the point
particle can hit the moving wall, say the one located at $\beta_+ \rightarrow -\infty$,
from anywhere within the wedge $|\beta_-| < -\sqrt{3} \beta_+$. As a result,
a few collisions are sufficient to make it so oblique that it will not bounce off
another wall. So, the universe will enter quickly in a definite Kasner trajectory and
stay there all the time in its approach to the singularity. Thus, the evolution
is not chaotic in these theories.

\section{Conclusions}
\setcounter{equation}{0}

We have shown that the homogeneous cosmologies provide consistent truncations of
Ho\v{r}ava-Lifshitz gravity in vacuum and investigated them using Hamiltonian
methods with emphasis on the closed space universe of Bianchi IX type. The
field equations reduce to an autonomous system of ordinary differential equations
describing the motion of a point particle in a potential well with $Z_3$ symmetry.
In general, the potential depends on time, through the volume moduli, but when
the universe approaches the initial singularity it freezes, as it becomes
independent of time. Then, for $\lambda > 1/3$, the universe flows to the
singularity by continuous changes of its shape, as in ordinary mixmaster cosmology,
rolling like a particle in the well with fixed (but arbitrary) energy $E$.
The main difference from general relativity is that the potential does not
vanish for generic values of the anisotropy parameters, and, thus, the evolution
of the early universe is not described by the Kasner solution away from the steep
walls. The dynamics is more intricate now, but, still, the shape of the potential
far away from the origin resembles that of general relativity and the particle can
bounce off the walls. In a certain limit (large $E$), the motion appears to be
non-ergodic, and, thus, chaos is absent. The same thing applies to very small values
of the energy $E$, though for a different reason. However, it remains to be seen
if the motion is chaotic for more general values of the parameter $E$,
as in general relativity, and compare it further with mixmaster dynamics in fully
covariant higher curvature generalizations of Einstein gravity. More work is
certainly required in this direction and we hope to return to it elsewhere.

This work should be considered as the beginning of a more general investigation
in Ho\v{r}ava-Lifshitz gravity. First, the most pressing open question is to revisit
the singularity theorems of general relativity and examine under which general
conditions they remain valid in theories with anisotropic scaling. In this context,
we will also be able to see whether the matter bounce of the Friedmann model
has more general value beyond the homogeneous and isotropic case. Assuming,
however, that the occurrence of space-like singularities is generic in
Ho\v{r}ava-Lifshitz gravity, it will be very important to examine how the
singularity is approached by extending the standard analysis of Belinskii,
Khalatnikov and Lifshitz. If the spatial points decouple
from the dynamics, which is a reasonable expectation even for theories with
anisotropic scaling in space and time, then, the homogeneous cosmologies
considered here will prove valuable tool for understanding the behavior of
the universe near the initial singularity.

Second, it is interesting to consider the canonical quantization of Bianchi IX
cosmology (or any other homogeneous model for that matter) as mini-superspace
models to Ho\v{r}ava-Lifshitz gravity. This can provide a tractable way to
compare it with Einstein gravity in the quantum regime. The Wheeler-DeWitt
equation for the "wave-function" of the universe appears to be more manageable
here because the walls of the effective potential are frozen in time in the domain
of validity of quantum cosmology. On the contrary, in ordinary quantum gravity,
the Bianchi IX model is more difficult to treat and interpret canonically because
the corresponding potential is not inert to the evolution, but scales with time.
We plan to address these issues in detail elsewhere.

Third, it should be noted that there is an intimate connection between the
Euclidean version of Ho\v{r}ava-Lifshitz gravity with "detailed balance"
and the theory of geometric flows. Namely, the gradient flow of the metric
derived from the functional (superpotential) ${\cal W}$ yields a continuous
deformation of the geometry on $\Sigma_3$ that is first order in time and trivially
satisfies the higher order equations of motion of the theory. This was first pointed
out in the original works, \cite{horava1}, \cite{horava2}, focussing, in
particular, in $2+1$ dimensional gravity with anisotropic scaling and its
connection to theory of Ricci flows on two-dimensional surfaces. In $3+1$
dimensions, the analogous deformation theory is provided by the so called
Cotton flow of
three-geometries, since the variation of the Chern-Simons term in the functional
${\cal W}$ is the Cotton tensor. This, then, provides the leading order term of the
flow, which is third order in space, and it should also be augmented with the
Ricci curvature terms that come about by varying the three-dimensional Einstein-Hilbert
term in ${\cal W}$. Remarkably, the Cotton flow admits consistent truncation
to an autonomous system of ordinary differential equations for all homogeneous
three-geometries, \cite{turk}, and the same thing applies
to the Ricci flow, \cite{jim}. In a forthcoming paper we discuss solutions
of the combined Cotton-Ricci flow for Bianchi IX geometries, which can be
thought as gravitational instanton solutions in the $3+1$ Ho\v{r}ava-Lifshitz
gravity, \cite{bblp}.
These configurations might also have important role in quantum gravity, in
the spirit of the Hartle-Hawking proposal for the construction of the
"wave-function" of the universe, using Euclidean path integrals. Thus, in
this context, it is natural to expect that the theory of geometric flows
will connect naturally to the problem of quantization of non-relativistic
theories of gravitation, in general, and for the Bianchi IX mini-superspace
model, in particular.

Finally, among other things, we mention the interesting possibility
of higher dimensional generalizations. In such cases, the theory is still
defined using the ADM decomposition of the metric, as in $3+1$ dimensions,
but the potential contains even higher curvature terms depending on the
space dimension. For example, in $4+1$ dimensions, we can have spatial
derivative terms up to order eight, which follow from a superpotential
${\cal W}$ that involves in its integrand the square of the Weyl tensor
$C_{ijkl} C^{ijkl}$ and the square of Ricci curvature $R^2$, assuming
the "detailed balance" condition, \cite{horava2}; one can also add the
corresponding Einstein-Hilbert term to ${\cal W}$, with cosmological
constant, to insure that the theory flows to five-dimensional Einstein
gravity at large scales . In such case, the Bach tensor of the four-dimensional
spatial geometry, which is obtained by varying the square of the Weyl tensor
in ${\cal W}$, replaces the Cotton tensor
that was featuring earlier in the $3+1$ dimensional theory; if an
$R^2$ term is also present in ${\cal W}$ one has to add its contribution,
which scales in the same way.
Then, the leading term of the potential $V_{\rm HL}$ (at least when the volume
of the five-dimensional universe is very small) is provided by the
square of the Bach tensor, plus possible additional contributions coming from
$R^2$ in ${\cal W}$, and it is scale invariant. In analogy with the
previous analysis, this term will dominate the cosmological evolution at early
times and lead to a frozen potential well, irrespective "detailed balance",
where the effective point
particle rolls. Similar considerations apply to all higher dimensional
cases. In view of the results obtained in the literature for higher
dimensional homogeneous string cosmology models, \cite{marc1},
\cite{marc2}, \cite{damour1}, \cite{damour2}, \cite{damour3}, it is also
important here to explore the universal features of dynamics close to the
singularity and expose their dependence on space-time dimensionality.

\vskip1cm

\section*{Acknowledgements}
This work was partially supported by the Cluster of Excellence "Origin and the
Structure of the Universe" in Munich, Germany, the French Agence Nationale pour
la Recherche programme "String Cosmology" under contract 05-BLAN-NT09-573739,
the ERC Advanced Grant "Mass Hierarchy and Particle Physics at the TeV Scale"
under contract 226371, the ITN programme "Unification in the LHC Era" under
contract PITN-GA-2009-237920 and the bilateral programme GRC APIC PICS-Gr\`ece
for French-Greek scientific collaboration under contract 3747. The authors I.B.,
F.B. and M.P. are grateful to the hospitality extended to them at the Arnold
Sommerfeld Center for Theoretical Physics in Munich during the early stages
of this work. We also thank Anastasios Petkou and Marika Taylor for stimulating
discussions during the GGI workshop "Perspectives in String Theory" that
triggered parts of the present work as well as Elias Kiritsis, Hermann Nicolai
and George Savvidy for fruitful exchanges. Finally, we thank Christos Sourdis
for assistance in drawing the figures.

\newpage

\appendix
\section{Bianchi IX model geometry}
\setcounter{equation}{0}

The Bianchi IX model describes a homogeneous (but generally non-isotropic)
three-dimensional geometry with the topology of $S^3$ and isometry group $SU(2)$.
The line element is constructed using the corresponding left-invariant 1-forms
$\sigma_i$,
\ba
\sigma_1 & = & {\rm sin} \psi {\rm sin} \theta d\phi + {\rm cos} \psi d\theta ~, \\
\sigma_2 & = & {\rm cos} \psi {\rm sin} \theta d\phi - {\rm sin} \psi d\theta ~, \\
\sigma_3 & = & {\rm cos} \theta d\phi + d\psi
\ea
and takes the form
\be
ds^2 = \gamma_1 \sigma_1^2 + \gamma_2 \sigma_2^2 + \gamma_3 \sigma_3^2 ~.
\ee

The 1-forms $\sigma_i$ satisfy the defining $SU(2)$ relations
\be
d\sigma_i + {1 \over 2} \epsilon_{ijk} \sigma_j \wedge \sigma_k  = 0 ~,
\ee
whereas the angles range as $0 \leq \theta \leq \pi$, $0 \leq \phi \leq 2\pi$ and
$0 \leq \psi \leq 4\pi$, since $\psi$ is extended to the double covering of the
rotation group. The space integration is carried out using
\be
\int \sigma_1 \wedge \sigma_2 \wedge \sigma_3 = \int {\rm sin} \theta ~ d\theta
\wedge d\phi \wedge d\psi = (4 \pi)^2 ~.
\ee

Next, we present the expressions for the Ricci curvature and Cotton tensors of
Bianchi IX metrics that will be used in the main text. Proper discussion requires
the use of $e^i = \sqrt{\gamma_i} ~ \sigma_i$ and the corresponding connection
1-forms ${\omega^i}_j$ satisfying the zero torsion relations
$de^i + {\omega^i}_j \wedge e^j = 0$. Then, the curvature 2-forms are computed as
\be
{R^i}_j = d{\omega^i}_j + {\omega^i}_k \wedge {\omega^k}_j
\ee
and the Ricci 1-forms are
\be
({\rm Ric})_i = i_k {R^k}_i ~.
\ee
The Ricci curvature scalar is $R = i_k ({\rm Ric})^k$. Also, the Cotton 2-form
is given by
\be
C^i = dY^i + {\omega^i}_j \wedge Y^j ~,
\ee
where $Y^j$ are simply given by
\be
Y^j = ({\rm Ric})^j - {1 \over 4} Re^j ~.
\ee

With these explanations in mind, the non-vanishing components of the Ricci
curvature tensor take the following form
\ba
R_{11} & = & {1 \over 2 \gamma_2 \gamma_3} \Big[\gamma_1^2 - (\gamma_2 - \gamma_3)^2
\Big] ~, \\
R_{22} & = & {1 \over 2 \gamma_1 \gamma_3} \Big[\gamma_2^2 - (\gamma_1 - \gamma_3)^2
\Big] ~, \\
R_{33} & = & {1 \over 2 \gamma_1 \gamma_2} \Big[\gamma_3^2 - (\gamma_1 - \gamma_2)^2
\Big]
\ea
and, therefore, the Ricci scalar curvature is
\be
R = -{1 \over 2 \gamma_1 \gamma_2 \gamma_3} \left(\gamma_1^2 + \gamma_2^2 + \gamma_3^2
- 2 \gamma_1 \gamma_2 - 2 \gamma_2 \gamma_3 - 2 \gamma_1 \gamma_3 \right) ~.
\ee
Also, the non-vanishing components of the Cotton tensor take the form
\ba
C_{11} & = & -{\gamma_1 \over 2 (\gamma_1 \gamma_2 \gamma_3)^{3/2}} \Big[\gamma_1^2
(2\gamma_1 - \gamma_2 - \gamma_3) - (\gamma_2 + \gamma_3) (\gamma_2 - \gamma_3)^2
\Big] ~, \\
C_{22} & = & -{\gamma_2 \over 2 (\gamma_1 \gamma_2 \gamma_3)^{3/2}} \Big[\gamma_2^2
(2\gamma_2 - \gamma_1 - \gamma_3) - (\gamma_1 + \gamma_3) (\gamma_1 - \gamma_3)^2
\Big] ~, \\
C_{33} & = & -{\gamma_3 \over 2 (\gamma_1 \gamma_2 \gamma_3)^{3/2}} \Big[\gamma_3^2
(2\gamma_3 - \gamma_1 - \gamma_2) - (\gamma_1 + \gamma_2) (\gamma_1 - \gamma_2)^2
\Big] ~.
\ea

It is convenient (and will be used throughout this paper) to parametrize the metric
coefficients $\gamma_i$ as follows,
\ba
\gamma_1 & = & e^{2 \Omega + \beta_+ + \sqrt{3} \beta_-} ~, \\
\gamma_2 & = & e^{2 \Omega + \beta_+ - \sqrt{3} \beta_-} ~, \\
\gamma_3 & = & e^{2 \Omega -2 \beta_+} ~.
\ea
The volume of $S^3$ is parametrized by $\Omega$, whereas $\beta_+$ and $\beta_-$ measure the
deviations from the isotropic metric that is associated to $\beta_+ = 0 = \beta_-$. Thus,
for $\gamma_1 \neq \gamma_2 \neq \gamma_3$, the metric on $S^3$ is homogeneous, but not
isotropic, having different circumference on great circles in each of the three mutually
orthogonal principal directions. Axially symmetric non-isotropic metrics are obtained by
choosing one of the deformation parameters equal to zero, say $\beta_- = 0$, in which case
$\gamma_1 = \gamma_2 \neq \gamma_3$. Likewise, $\gamma_1 \neq \gamma_2 = \gamma_3$
requires $\beta_- = \sqrt{3} ~ \beta_+$ and $\gamma_1 = \gamma_3 \neq \gamma_2$ requires
$\beta_- = - \sqrt{3} ~ \beta_+$.

Finally, note that the Cotton tensor vanishes in the isotropic case $\gamma_1 = \gamma_2
= \gamma_3$, since the metric of the round $S^3$ is conformally flat.

\section{The bounce law}
\setcounter{equation}{0}

In this appendix, we derive the bounce law of the effective point particle as it
hits the wall located at $\beta \rightarrow - \infty$. Far away from the wall the
potential is assumed to vanish and the particle moves freely, where this is
applicable.

{\bf General relativity:} In this case, the Hamiltonian is derived using the
asymptotic form of the potential $V_{\rm GR}$, as $\Omega \rightarrow - \infty$,
and becomes approximately
\be
2H_{\rm GR} = p_+^2 + p_-^2 - {1 \over 4} p_{\Omega}^2 + {12 \over \kappa^4}
e^{4(\Omega - \beta_+)} ~.
\ee

The $\Omega$-dependence of the potential can be easily transformed away by
introducing new variables
\be
\bar{\Omega} = {1 \over \sqrt{3}} \left(2 \Omega - {\beta_+ \over 2} \right) ~,
~~~~~~
\bar{\beta}_+ = {1 \over \sqrt{3}} (\beta_+ - \Omega)
\ee
and their conjugate momenta
\be
\bar{p}_{\Omega} = {2 \over \sqrt{3}} (p_{\Omega} + p_+) ~, ~~~~~~
\bar{p}_+ = {1 \over \sqrt{3}} (p_{\Omega} + 4 p_+) ~.
\ee
Then, the Hamiltonian takes the simpler form
\be
2H_{\rm GR} = {1 \over 4} \bar{p}_+^2 + p_-^2 - {1 \over 4} \bar{p}_{\Omega}^2 +
{12 \over \kappa^4} e^{- 4 \sqrt{3} \bar{\beta}_+} ~.
\label{newvari}
\ee
In terms of these variables, the potential is frozen in time and both
$\bar{p}_{\Omega}$ and $p_-$ are constants of motion. Dividing by parts, one obtains
\be
{\bar{p}_{\Omega} \over p_-} = {2 \over \sqrt{3}} {1 + p_+ / p_{\Omega} \over
p_- / p_{\Omega}} = {\rm constant}
\label{laue}
\ee
relating the original momenta before and after the bounce.

At this point, it is useful to introduce Kasner exponents, which are applicable
to the evolution of the universe well before and after the bounce,
\ba
& & n_1 = {1 \over 3p_{\Omega}} (p_{\Omega} - 2 p_+ - 2 \sqrt{3} p_-) ~, ~~~~~
n_2 = {1 \over 3p_{\Omega}} (p_{\Omega} - 2 p_+ + 2 \sqrt{3} p_-) ~, \nonumber\\
& & n_3 = {1 \over 3p_{\Omega}} (p_{\Omega} + 4 p_+)  ~,
\ea
and satisfy $n_1^2 + n_2^2 + n_3^2 = 1$ by virtue of the Hamiltonian constraint
$p_{\Omega}^2 = 4(p_+^2 + p_-^2)$ far away from the wall. It is also convenient to
parametrize the Kasner exponents in terms of a single variable $s$, following Misner,
\cite{misner1},
\be
n_1 = {2s (s-3) \over 3 (s^2 + 3)} ~, ~~~~~
n_2 = {2s (s+3) \over 3 (s^2 + 3)} ~, ~~~~~
n_3 = -{(s-3)(s+3) \over 3 (s^2 + 3)} ~,
\label{thepara}
\ee
so that the constant of motion \eqn{laue} takes the form
\be
{\bar{p}_{\Omega} \over p_-} = 2 {n_3 + 1 \over n_2 - n_1} = {s \over 3} +
{3 \over s}  = {\rm constant} ~.
\ee

Thus, in general relativity, the bounce law against the wall $\beta_+ \rightarrow -
\infty$ is simply described as
\be
{s \over 3} \rightarrow {3 \over s} ~,
\ee
which leaves $\bar{p}_{\Omega} / p_-$ invariant and changes the Kasner exponents
accordingly.

{\bf Ho\v{r}ava-Lifshitz gravity:} The Hamiltonian is now derived using the
asymptotic form of the potential $V_{\rm HL}$, which is independent of $\Omega$
when $\Omega \rightarrow - \infty$, and becomes approximately
\be
2H_{\rm HL} = p_+^2 + p_-^2 - {1 \over 2(3 \lambda - 1)} p_{\Omega}^2 +
{18 \alpha  \over \kappa^2} e^{-12 \beta_+} ~.
\ee
The point particle moves freely before and after the bounce only when
$p_{\Omega}$ is very large, in which case the Hamiltonian constraint simplifies to
$p_{\Omega}^2 = 2 (3\lambda - 1) (p_+^2 + p_-^2)$ for generic values
of $\beta_{\pm}$. This will be implicitly assumed here and also that $\alpha > 0$
and $\lambda > 1/3$.

In the present case, there is no $\Omega$-dependence on the wall located at $\beta_+
\rightarrow - \infty$ and it follows immediately that
\be
{p_{\Omega} \over p_-} = {\rm constant} ~.
\ee
Then, this is an ordinary bounce from a steady wall following the standard
rule that the incidence and reflection angles are equal asymptotically,
namely $p_+$ flips sign and $p_-$ remains unchanged. To compare with the
previous case, it is convenient to
describe the free motion of the particle well before and after the bounce
using Kasner exponents, which are now defined as
\ba
& & n_1 = {1 \over 3p_{\Omega}} \left(p_{\Omega} - \sqrt{2(3\lambda -1)} p_+ -
\sqrt{6(3\lambda -1)} p_- \right) ~, \nonumber\\
& & n_2 = {1 \over 3p_{\Omega}} \left(p_{\Omega} - \sqrt{2(3\lambda -1)} p_+ +
\sqrt{6(3\lambda -1)} p_- \right) ~, \nonumber\\
& & n_3 = {1 \over 3p_{\Omega}} \left(p_{\Omega} + 2\sqrt{2(3\lambda -1)} p_+
\right)
\ea
and satisfy $n_1^2 + n_2^2 + n_3^2 = 1$ by virtue of the Hamiltonian constraint
far away from the wall. By employing Misner's parametrization \eqn{thepara}, as
before, we obtain
\be
{p_{\Omega} \over p_-} = \sqrt{{2(3\lambda -1) \over 3}} {2 \over n_2 - n_1} =
\sqrt{{3 \lambda -1 \over 2}} \left({s \over \sqrt{3}} + {\sqrt{3} \over s}
\right) = {\rm constant} ~.
\ee

Thus, the bounce law against the wall at $\beta_+ \rightarrow - \infty$ is now
described as
\be
{s \over \sqrt{3}} \rightarrow {\sqrt{3} \over s} ~,
\ee
which leaves $p_{\Omega}/p_-$ invariant and changes the Kasner exponents accordingly.
It has the same from as in general relativity, $s/3 \rightarrow 3/s$, setting
\be
s_{\rm GR} = \sqrt{3} ~ s_{\rm HL} ~.
\ee

Summarizing, the bounce in Ho\v{r}ava-Lifshitz gravity follows the standard
reflection rule from a steady wall, whereas in general relativity this
rule is modified by the moving walls and it is effectively described by inserting a
factor of $\sqrt{3}$ in the corresponding Misner parameter. In general relativity,
the standard rule of equal incidence and reflection angles only applies to the
transformed Hamiltonian \eqn{newvari} from which the bounce law for the original
momenta $p_{\pm}$ was derived.

\end{document}